\documentclass[a4paper]{article}

\usepackage{comment}
\usepackage{eurosym}
\usepackage{amsfonts}
\usepackage{graphicx}
\usepackage{amsmath}
\usepackage{flafter}
\usepackage{subfloat}
\usepackage{subfigure}
\usepackage{fancyhdr}
\usepackage{color}
\usepackage{flafter}
\usepackage{mathabx}
\usepackage[numbers,sort&compress]{natbib}
\usepackage{caption}

\begin{document}

\date{}
\title{Digital clocks: simple Boolean models can quantitatively describe
circadian systems}
\author{Ozgur E. Akman$^{1,3,\asterisk}$, Steven Watterson$^{2,3}$, Andrew
Parton$^{3,4}$, \\
Nigel Binns$^{2}$, Andrew J. Millar$^{3,\dagger}$ and Peter Ghazal$%
^{2,3,\dagger}$ }
\maketitle

\noindent
$^{1}$Centre for Systems, Dynamics and Control, College of Engineering, Computing and
Mathematics, University of Exeter, Harrison Building, North Park Road,
Exeter EX4 4QF, United Kingdom.\\
\noindent
$^2$Division of Pathway Medicine, University of Edinburgh Medical,
Chancellors Building, 49 Little France Crescent, Edinburgh EH16 4SB, United Kingdom.\\
\noindent
$^3$SynthSys Edinburgh, University of Edinburgh,
CH Waddington Building, Kings Buildings, Mayfield Road, Edinburgh EH9 3JD, United Kingdom.\\
\noindent
$^4$Department of Mathematics, University of Edinburgh, James Clerk
Maxwell Building, Kings Buildings, Mayfield Road, Edinburgh EH9 3JZ, United Kingdom.\\

\noindent{$^{\asterisk}$Corresponding author. E-mail:
O.E.Akman@ex.ac.uk; Tel: +44 1392 724 060; Fax: +44 1392 217 965.}

\noindent{$^{\dagger}$These two authors are the joint senior authors of
this paper.}

\bigskip \noindent{Keywords: systems biology; circadian gene networks;
Boolean logic; photoperiodism; \textit{Arabidopsis thaliana}}

\bigskip \noindent{Running title: Digital clocks.}

\section*{Abstract}

The gene networks that comprise the circadian clock modulate biological function across a range of scales, from gene expression to performance and adaptive behaviour. The clock functions by generating endogenous rhythms that can be entrained to the external 24-hour day/night cycle, enabling organisms to optimally time biochemical processes relative to dawn and dusk. In recent years, computational models based on differential equations have become useful tools for dissecting and quantifying the complex regulatory relationships underlying the clock's oscillatory dynamics. However, optimising the large parameter sets characteristic of these models places intense demands on both computational and experimental resources, limiting the scope of \textit{in silico} studies.

Here, we develop an approach based on Boolean logic that dramatically reduces the parametrisation, making the state and parameter spaces finite and tractable. We introduce efficient methods for fitting Boolean models to molecular data, successfully demonstrating their application to synthetic time courses generated by a number of established clock models, as well as experimental expression levels measured using luciferase imaging. Our results indicate that despite their relative simplicity, logic models can: (i) simulate circadian oscillations with the correct, experimentally-observed phase relationships amongst genes; and (ii) flexibly entrain to light stimuli, reproducing the complex responses to variations in daylength generated by more detailed differential equation formulations. Our work also demonstrates that logic models have sufficient predictive power to identify optimal regulatory structures from experimental data. 

By presenting the first Boolean models of circadian circuits together with general techniques for their optimisation, we hope to establish a new framework for the systematic modelling of more complex clocks, as well as other circuits with different qualitative dynamics. In particular, we anticipate that the ability of logic models to provide a computationally efficient representation of system behaviour could greatly facilitate the reverse-engineering of large-scale biochemical networks.

\section{Introduction}

Circadian rhythms are the fundamental daily oscillations in metabolism,
physiology and behaviour that occur in almost all organisms, ranging from
cyanobacteria to humans \citep{Dunlap03}. The gene regulatory networks
(GRNs), or \textit{clocks}, that generate these rhythms regulate the
expression of associated genes in roughly 24-hour cycles. Circadian networks
have been studied in a variety of experimentally tractable model systems,
revealing that different organisms share structurally similar circuits based
around interlocking sets of positive and negative gene-protein feedback
loops \citep{Young01,Bell-Pedersen05,Zhang10}. In mammals, circadian rhythms
are being increasingly recognised as important to healthy phenotypes, playing a
role in ageing \citep{Khapre10}, cancer \citep{Gery10}, vascular disease %
\citep{Takeda10} and psychiatric disorders \citep{Westrich10}, as well as
modulating innate immunity \citep{Lange06,Liu06,Keller09}.

Clocks synchronise to their environment by using light and temperature to
regulate the levels of one or more components of the feedback loops. This
ensures that key biological processes are optimised relative to dawn and
dusk, benefiting growth and survival \citep{Ouyang98,Dodd05a}. For the
clock to provide such an adaptive advantage, the phase must change
appropriately when the clock is subject to regular perturbations - particularly
seasonal changes in daylength (the \textit{photoperiod}). However, as well as exhibiting flexible responses to  variations in the input light signal, the clock must also exhibit robustness to irregular perturbations, such as genetic mutations and the intrinsically stochastic environment of the cell. 

Temperature also plays a critical role as an environmental time cue. Across different species, the clock is relatively insensitive to temperature in that the period of free-running oscillations typically has a $Q_{10}$ value close to 1 \citep{Gardner81,Ruoff97,Gould06}. This latter phenomenon, known as temperature compensation, is generally considered to be one of the defining properties of the circadian clock and has been suggested to be a key requirement for stability of the clock's phase relationship under seasonal temperature variations \citep{Brunner06,Akman08}.

The ability of ordinary and delay differential equations (\textit{DE}s) to
reproduce the underlying continuous dynamics of biochemical networks, and to
parametrise individual reactions, has led to the construction of DE models
in a number of circadian organisms. These include the fungus \textit{%
Neurospora crassa} %
\citep{Leloup99,Smolen01,Sriram04,Francois05,Ruoff05,Hong08a,Akman10}, the
fly \textit{Drosophila melanogaster} %
\citep{Goldbeter95,Leloup98,Ueda01,Smolen04}, the mammal \textit{Mus musculus%
} \citep{Forger03,Leloup03,Becker04,Mirsky09} and the higher plant \textit{%
Arabidopsis thaliana} %
\citep{Locke05a,Locke05b,Locke06,Zeilinger06,Pokhilko10}. Such models have
proven useful in uncovering the general design principles of circadian
oscillators, as well as providing a quantitative framework within which to
interpret experimental results \citep{Pokhilko10,Zhang10}. In particular,
novel insights have been gained into the mechanisms promoting robustness
with respect to photoperiod changes \citep{Akman10}, temperature
fluctuations \citep{Ruoff05,Akman08} and molecular noise  %
\citep{Gonze02a,Gonze02b,Vilar02}. The DE models have also yielded experimentally-testable predictions that have lead
to the discovery of novel circadian regulators \citep{Locke06}.

However, a significant drawback of the DE approach is that the values of the
kinetic parameters controlling each individual reaction have to be
specified, and for clocks these are typically unknown. When constructing a
DE\ model, it is therefore necessary to calculate the particular combination
of parameter values giving an optimal fit to experimental data %
\citep{Locke05a,Locke05b,Locke06,Zeilinger06,Akman08,Akman10}. For realistic
systems involving large numbers of reactions, this optimisation procedure is
computationally very expensive making exhaustive parameter searches intractable. With increasing parameter numbers also comes a
need for data with which to constrain the optimisation, placing a greater
demand on experiment in terms of finance, time and ethics. These concerns mean that there is a pressing need for modelling approaches
that minimise the number of parameters required, whilst adequately capturing
the essential dynamical behaviour of the system of interest.

Here we develop just such an approach, based on Boolean logic. In Boolean
models, the activity of each gene is described with a two-state variable
taking the value ON (1) or OFF (0), meaning that its products are present or
absent respectively. Biochemical interactions are represented by simple,
binary functions which calculate the state of a gene from the activation
state of its upstream components %
\citep{Kauffman69,Thomas91,Kaufman99,Shmulevich02a,Watterson08a,Watterson08b,Saez09,Schlatter09,Watterson10}%
. This approximation dramatically reduces the state-space of the system,
mapping the infinite number of different continuous system states in a DE
model to a finite number of discrete states in the Boolean equivalent.

An additional important advantage of using a logic approach is that the total number of
parameters is substantially reduced. For a given gene, the full set of
reactions determining its state through a particular interaction is
parametrised by a single \textit{signalling delay}, representing the net
time taken for these reactions to cause a change in state. Fitting to
experimental data introduces an associated \textit{discretisation threshold}%
, with expression levels above the threshold taken to correspond to the ON
state of the gene and levels below it to the OFF state %
\citep{Watterson08a,Watterson08b,Saez09,Watterson10}. In fitting a
particular experimental data set, each delay becomes a multiple of the
sampling interval, while only a bounded subset of thresholds will yield
distinct Boolean expression patterns. This means that the total number of
parameter combinations is finite and can be enumerated. Thus, by building a
logic version of a DE model, an infinite model can be converted into a finite
one with fewer parameters to be optimised. This extends the scale and
complexity of GRNs that can be studied by Boolean models far beyond the practical scope of DEs.

In this work, we introduce the first Boolean models of circadian networks.
By constructing logic analogs of a number of established DE\ clock models,
we demonstrate that in each case the Boolean models are capable of
accurately reproducing the higher-order properties - particularly
photoperiod responses - of their DE counterparts. This suggests that the
complex, biological signal transduction simulated by the DE models can be
captured in Boolean equivalents possessing significantly smaller parameter
sets. We introduce a general method for optimising Boolean models that avoids the
qualitative and often subjective terms characteristic of the cost functions
used to fit the parameters of large DE clock models. Furthermore, we
show that our fitting algorithm is capable of determining the optimal
Boolean model configuration associated with a given circuit topology and
experimental data set. In particular, our algorithm successfully predicts the recently discovered 
repressive action of the circadian gene TOC1 on LHY in the central feedback loop of the \textit{Arabidopsis} clock \citep{Gendron12}.

Taken together, our results show that Boolean models can quantitatively distinguish between
a range of putative regulatory structures on the basis of the system
dynamics. This identifies Boolean logic as a viable technique for reverse
engineering circadian networks, complementing approaches based on DEs.
Moreover, our work also suggests novel hybrid modelling approaches based on employing Boolean models as a first step towards
the construction of more detailed DE formulations. More generally, we propose that our methodology provides an efficient way of systematically modelling complex signalling pathways, including other oscillatory circuits and systems characterised
by steady-state dynamics.

\section{Results}

\subsection{Logic models employ significantly fewer parameters}

We selected 4 recent circadian oscillator models of increasing complexity with which to assess the suitability of a Boolean formulation. The simplest of these was a \textit{Neurospora} model based on a
single negative feedback loop with a single light input \citep{Leloup99}
(Figure \ref{fig:DECircuitDiags}A). This is represented by 3 differential
equations (\textit{DE}s) parametrised by 13 kinetic constants. The second
model was a modified version of the 1-loop \textit{Neurospora} circuit in
which there are a pair of negative feedback loops associated with different
isoforms of the active protein \citep{Akman08} (Fig. \ref{fig:DECircuitDiags}B). The extra feedback loop results in 5 DEs parametrised by 18 kinetic
constants. The third model was an \textit{Arabidopsis} circuit based on a
pair of interlocking feedback loops with 3 light inputs \citep{Locke05b}
(Fig. \ref{fig:DECircuitDiags}C). It is described by 13 DEs together with 64
kinetic constants. The final model considered was a 3-loop \textit{%
Arabidopsis} circuit obtained by adding an extra feedback loop and light
input to the 2-loop system \citep{Locke06} (Fig. \ref{fig:DECircuitDiags}D). This yields 16 DEs parametrised
by 80 kinetic constants.

In a Boolean model, an interaction $j$ between two components $X_{i}$ and $%
X_{k}$ is quantified by the corresponding signalling delay $\tau _{j}$. This
is the time taken for the biochemical processes represented by $j$ to
convert a change in the state of $X_{i}$ into a change in the state of $%
X_{k} $ (see Fig. S1 of the Electronic Supplementary Material). The signalling delays are thus parameters that
determine the dynamics of the model, with different combinations of delays
yielding different attractor states (e.g. steady-states or limit cycles) %
\citep{Thomas83,Kaufman85,Thomas91,Kaufman99}. In order to relate the
discrete dynamics to the continuous variations in expression level observed
experimentally, it is necessary to introduce a discretisation process. Supplementary Fig. S2A shows a hypothetical time course and Supp. Fig. S2B the
result of applying a particular discretisation threshold. In the general
case, the choice of threshold is dependent on the effective range of
activity and expression that is critical for signal propagation. As a
result, each threshold also becomes a parameter of the logic model %
\citep{Watterson08a,Watterson08b,Watterson10}.

For logic descriptions of the circadian networks shown in Fig. \ref{fig:LCDiags}, each edge is parametrised by a signalling delay $\tau _{j}$
and each vertex by a threshold $T_{i}$. Thus, a network is characterised by
the vectors $\mathbf{\tau }=\left( \tau _{j}\right) $ and $\mathbf{T}=\left(
T_{i}\right) $. In Fig. \ref{fig:hist_param}, we compare the total number of
parameters in the logic and DE versions of each network (see also
Supplementary Table S1). We can see that dramatically fewer
parameters are required in a logic description compared to the corresponding DE
formulation. Indeed for the largest network considered, 3-loop \textit{Arabidopsis}, the Boolean description reduces the number of parameters by a factor of 4. 

\subsection{Logic model configurations consistent with DE models are optimal%
}

In identifying the logic models that best reproduce the corresponding DE
dynamics, we considered variations to the structures of the networks shown
in Fig. \ref{fig:DECircuitDiags}. Starting from the abstract topologies of Fig. \ref{fig:LCDiags}, 
each edge is activating or inhibiting,
and where two or more edges lead into a vertex, the corresponding
inhibition/activation signals are combined to determine the state of the
gene.

In the Boolean formalism, the manner in which regulatory signals affect a
gene's expression level is represented by the corresponding \textit{logic
gate}. This is a binary function that specifies the current state of the
gene, ON (1) or OFF (0), for each possible combination of input states \citep%
{Shmulevich02b,Watterson08a}. For genes with a single input, there are two
possible functions for which the output varies with the input: the identity gate, in which the output follows the input
($0\rightarrow 0$ and $1\rightarrow 1$), and the NOT gate, in which the
input is inverted ($0\rightarrow 1$ and $1\rightarrow 0$). These represent
activation and repression of the gene by its regulator respectively. Genes
with two inputs are commonly modelled using either an AND\ or an OR gate %
\citep{Shmulevich02b,Watterson08a,Schlatter09}. For simplicity, we considered each
multi-input gate to be a composition of ANDs and ORs (see Methods for
further details). The particular combination of logic gates used to model a
network is referred to here as its \textit{logic configuration} (%
\textit{LC}): this encodes the regulatory structure of the network in a
compact fashion.

It follows that each abstract topology of Fig. \ref{fig:LCDiags} gives
rise to $2^{E+M}$ possible LCs, where $E$ is the number of edges and $M$ is
the number of vertices with more than one input. The total number of
possible LCs is 4 for 1-loop \textit{Neurospora}, 32 for 2-loop \textit{%
Neurospora}, 256 for 2-loop \textit{Arabidopsis} and 2048 for 3-loop \textit{%
Arabidopsis}. In each case, a subset of these LCs are consistent with the
pattern of activation and inhibition in the corresponding DE model. That
there can be more than one such LC in each case is due to the choice of AND or OR
where a vertex has multiple edges leading into it. For \textit{Neurospora},
the 1-loop circuit has a unique LC consistent with the DE model while for
the 2-loop circuit there are two DE LCs. The 2- and 3-loop \textit{%
Arabidopsis} networks yield 4 and 8 DE LCs respectively.

For a given logic model, we would expect the DE\ LCs to most closely
reproduce the DE dynamics. To test this hypothesis, we optimised LCs to
synthetic experimental data obtained from the DE system, and then compared
their predictive performance. For each LC, the match to the continuous
dynamics was quantified by finding the combinations of parameters
(signalling delays and discretisation thresholds) that minimised a quantitative\
cost function. In order to be able to objectively compare the ability of the Boolean models to reproduce experimental time courses against that of their DE counterparts, we employed a cost function that closely mirrored those commonly used to optimise continuous models \citep%
{Locke05a,Locke05b,Locke06,Akman08,Akman10,Pokhilko10}. The cost function we used measured the goodness-of-fit of each logic
model to synthetic data generated in both 24h light-dark (LD) cycles and the
appropriate free-running light regime (continuous dark, DD, for \textit{%
Neurospora};\textit{\ }continuous light, LL, for
\textit{Arabidopsis}). At each vertex, the
cost score was calculated as the correlation between the discretised time
series for the downstream species and the time-delayed predicted output
calculated from the discretised data for the upstream species. Scores were
summed across all vertices and light regimes to give the final cost value
(see Methods for details).

After ranking the optimised LCs by score, we then assessed whether the top
ranking LCs comprised viable clock circuits by checking that they were
capable of: (i) generating self-sustained oscillations with a circadian
period in constant conditions; and (ii) entraining to LD cycles over a
realistic range of photoperiods \citep{Troein10}.

For the \textit{Neurospora} and 2-loop \textit{Arabidopsis }logic models,
the full set of LCs were fitted to synthetic data. The best-performing LCs
for these networks are shown in Figs. \ref{fig:NeuroScores} and \ref{fig:ArabScores}A. It can be seen that the optimisation method produces a
clear separation of the LCs by score. Moreover, for each network, one of the
DE LCs is uniquely identified as the optimal circuit yielding a viable
clock. In the case of 3-loop \textit{Arabidopsis}, we considered
the subset of configurations obtained by setting the gates common with the
2-loop \textit{Arabidopsis} circuit to their optimised values. This mirrored
the construction of the 3-loop DE\ model which was derived from the 2-loop
system by adding an additional feedback loop while fixing all other
interactions, and then optimising the parameters of the new loop %
\citep{Locke06}. Constraining the structure of the circuit in this fashion
yielded 8 possible LCs (the 2 edges in the LHY-PRR loop described as
activation or inhibition and the AND/OR interaction at LHY; see Fig. \ref{fig:LCDiags}D). Figure \ref{fig:ArabScores}B shows that in this system, as
for the other models, a DE LC emerges as the optimal circuit.

\subsection{Optimal Boolean models have biological time series
characteristics}

The time series generated by the optimal configurations in LD cycles are
shown in Fig. \ref{fig:LDts}. The corresponding DE simulations are also
plotted for comparison.

In each case it is clear that the Boolean models capture the same
qualitative dynamics as their DE counterparts. Different species are
switched on and off relative to one another with phases that match the
patterns of rising and falling expression in the corresponding continuous
time series. Moreover, the delays between the switching times are similar to
the phase differences between the peaks and troughs of the DE solutions.

It should be noted that both Boolean \textit{Arabidopsis}
models reproduce the acute light response in the Y gene, as well as the
circadian response in Y around dusk (cf. Figs. \ref{fig:LDts}E-H). This
demonstrates the ability of the Boolean circuits to simulate biochemical
processes that occur on different time scales within the same system.

The optimal LCs give equally good matches to the DE dynamics in simulated
free-running conditions, as can be seen in Supp. Fig. S3.

\subsection{Optimal Boolean models have biological photoperiodic behaviour}

In order to assess the extent to which the Boolean models reproduce the DE
dynamics in a more global and quantitative fashion we compared phase
changes over a range of biologically realistic photoperiods. In the Boolean
framework, the times at which solutions switch between 0 and 1 emerge as
natural phase measures. For continuous time courses - such as those generated
by DE models - the point in the circadian cycle at which the expression
level of a molecular species decreases below a set threshold can be employed as a phase marker \citep{Tan04,Akman10,Edwards10}. This suggested using the time at which
each species decreases below its discretisation threshold as a phase measure
for the DE simulations, and the time of the $1\rightarrow 0$ transition as
the equivalent marker in the Boolean models.

The phase-photoperiod relationships computed in this fashion are shown for
the \textit{Neurospora} models in Figs. \ref{fig:PPs}A and B. For both
networks, the photoperiodic behaviour of the Boolean and DE models is very
close: indeed for the 1-loop network, they are exactly equivalent. 
Figs. \ref{fig:PPs}C and D plot the photoperiod simulations obtained with the \textit{%
Arabidopsis} circuits. Here too, the phase-photoperiod profiles are very
similar, with the addition of the LHY-PRR\ loop to the 2-loop model causing a transition from a
predominately dusk-locked system to a dawn-locked one \citep{Edwards10}. In
particular, the Boolean 2-loop \textit{Arabidopsis} circuit exactly
reproduces the dual light response in the Y gene, in which the acute peak
tracks dawn and the circadian peak tracks dusk. This suggests that the logic
circuits possess sufficient dynamic flexibility to perform the complex
integration of environmental signals that is a central property of circadian systems.

\subsection{Boolean models can determine circadian network structure from
experimental data}

The success of the logic models in recovering the correct DE configurations
from synthetic data suggested that for a fixed abstract topology, our
optimisation procedure has the capacity to determine the logic network most
consistent with a given data set. We tested this finding further by
optimising the 3-loop \textit{Arabidopsis }logic circuit to highly-sampled
experimental time series recorded using luciferase (LUC)\ imaging in
constant light from a wild-type strain \citep{Edwards10}. All possible LCs were considered,
corresponding to a network inference carried out assuming no prior
biological knowledge. The cost function optimised was the same as that used
for fitting to synthetic LL data. As previously, viable clock circuits were
taken to be those yielding autonomous limit cycles with circadian periods.

The results of fitting to experimental data are presented in Fig. \ref{fig:ArabExpScores}A. 
It can be seen that the second highest ranking LC
giving a viable circuit is a DE configuration. This
configuration, $G_{DE}$, is in fact the same as that previously determined
to be optimal from the synthetic \textit{Arabidopsis} data sets. Moreover, Fig. \ref{fig:ArabExpScores}B shows that $%
G_{DE}$ emerges as the top ranking clock configuration if the regulatory structure is constrained to incorporate the central LHY-TOC1 negative feedback loop of the corresponding continuous model.
Interestingly, the optimal configuration, $G_{OPT}$, differs from $G_{DE}$ in the sign of the TOC1-LHY interaction, with TOC1 repressing LHY in $G_{OPT}$ as opposed to activating it (compare Supp. Figs. S4A and B). This result is consistent with the recent experimental characterisation of the LHY-TOC1 circuit as a double negative feedback loop, rather than the single negative loop assumed in the 2- and 3-loop DE models \citep{Gendron12}.

Figures \ref{fig:expfreerunts}B and C show the simulations of the
free-running clock obtained from the optimal and DE configurations
respectively. For both LCs, the simulated oscillation period and timing of gene expression is close to that of
the natural system (cf. Fig. \ref{fig:expfreerunts}A). This can be seen clearly by comparing the durations for which each gene is ON in the two Boolean models with the corresponding peaks in the continuous data.

\section{Discussion}

\subsection{A new approach to quantitative circadian modelling based on Boolean logic}

Circadian clocks have become popular systems for studying the relationship
between gene-protein dynamics and phenotype. The molecular machinery
underlying these networks has been relatively well characterised, and this
has led to great interest in developing predictive computational models of
the clock. Such models are usually formulated as sets of differential
equations. The high level of biochemical detail afforded by this approach
has allowed DE models to successfully address a range of issues
regarding the functional relationship between the architecture of the clock
and circadian homeostasis. One homeostatic mechanism that has been comprehensively investigated from a theoretical perspective is temperature compensation. By assuming that certain subsets of a circuit's kinetic parameters are temperature-dependent, temperature has been incorporated into a number of circadian DE models, ranging from minimal circuits built on the Goodwin oscillator \citep{Ruoff96,Ruoff97,Ruoff05} to more detailed formulations that explicitly model the underlying biochemical reactions \citep{Akman08,Hong08a}. These models have provided new insights into the possible mechanisms that lead to circadian mutants with altered compensation properties, as well as suggesting generic motifs that may facilitate the tuning of the phase-temperature relationship.

However, clock models based on DEs are characterised by large numbers of
kinetic parameters, the values of which are typically unknown. Optimising
these to experimental data is a major computational bottleneck which imposes
a hard bound on the maximum system size that can be studied. In addition, as
the fitting problem is typically highly underdetermined and involves noisy,
undersampled experimental data, robust optimisation can require the
construction of cost functions targeting specific qualitative features of
the data, introducing a degree of arbitrariness to the fitting procedure %
\citep{Locke05a,Locke05b,Locke06,Akman08,Akman10}.

The need for minimal clock parametrisations to address these issues has been
recognised elsewhere in the literature. For example, recent \textit{%
Neurospora} work has shown that it is possible to use a simple two-parameter
function to represent all the intermediate processes between the expression
of a clock gene and its action on a downstream target, whilst still
maintaining sufficient flexibility to accurately simulate biological
temperature and photoperiod responses \citep{Akman08,Akman10}.

An alternative technique for modelling gene regulatory networks is provided
by Boolean logic. By assuming discrete expression levels, Boolean models
provide an even greater reduction in complexity, albeit at the cost of
reduced biochemical precision. Previous studies have exploited this reduction to study GRN state-space
structures \citep{Ay09,Saez09,Hau11}, and to introduce probabilistic models
that parametrise statistical transitions between states %
\citep{Shmulevich02a, Krawitz07a}. The different limit cycles enumerable by
a single Boolean GRN have been interpreted in some models as examples of
cell differentiation \citep{Ribeiro07, Garg08}, making limit cycle
properties of special interest. Elsewhere, Boolean studies have focused on the regulatory logic underlying transcription \citep{Buchler03,Gerstung09,Hunziker10}, the immune response %
\citep{Kaufman85,Kaufman99} and apoptosis \citep{Schlatter09}.

Here, we have presented the first models of
biological clocks based on the Boolean formalism. We constructed logic
versions of 4 DE\ models simulating circadian rhythms in the organisms \textit{Neurospora crassa} and \textit{Arabidopsis thaliana}. To
derive the best logic representations, both regulatory structure (logic
configuration) and parameters (signalling delays and discretisation
thresholds) were optimised to synthetic experimental data generated from the
DEs. In each case, we found that the logic configuration yielding the best
fit to data was consistent with that of the corresponding DE system. This
suggested that the Boolean models were capable of identifying the particular
combination of activating and inhibiting elements best able to account for a
set of experimental expression patterns (Figs. \ref{fig:NeuroScores} and \ref{fig:ArabScores}). We confirmed this hypothesis by successfully applying our
optimisation algorithm to real \textit{Arabidopsis} LUC\ data (Fig. \ref{fig:ArabExpScores}). These results demonstrate that the
logic models possess sufficient predictive power to perform biological
network inference, despite their relative simplicity. Moreover, the optimisation to LUC data highlights the importance of network structure in determining dynamic behaviour \cite{Milo02,Prill05}. This optimisation gave rise to a pair of putative 3-loop network architectures with distinct parameter values that generated very similar expression time series (Supp. Fig. S4 and Fig. \ref{fig:expfreerunts}). One of these architectures matched the 3-loop \textit{Arabidopsis} DE model. It therefore contains the DE model's core LHY-TOC1 negative feedback loop, which was based on an experimental study that demonstrated the repression of TOC1 by LHY whilst also inferring the activation of LHY by TOC1 \citep{Alabadi01}. Significantly, the other architecture - which gave the best fit to data - agreed with more recent biochemical work showing that TOC1 is in fact a repressor of LHY \citep{Gendron12}. This result is also consistent with a complementary computational study showing that an expanded version of the 3-loop DE model incorporating a negative TOC1-LHY interaction yields more accurate simulations of TOC1 knockout and overexpression mutants \citep{Pokhilko12}. 

We also found that although the cost function used to fit synthetic data only assessed goodness-of-fit in simulated 12:12 LD cycles, the logic models
gave very good fits to the DEs in long and short days also, closely
reproducing the relationships between photoperiod and expression phase (Fig. \ref{fig:PPs}). The coordination of biochemical activity with the timing of
dawn and dusk is a key system-level property that can be used to assess the
relationship between the structure of a circadian network and its
evolutionary flexibility \citep{Edwards10}. Our photoperiod simulations
indicate, somewhat surprisingly, that this property can be accounted for by
significantly reduced dynamic models possessing much smaller parameter sets.
In particular, the combined dawn- and dusk-tracking observed in the 2-loop 
\textit{Arabidopsis} DE model (Fig. \ref{fig:PPs}C) is accurately replicated
by its logic formulation which has less than 1/4 the number of parameters. A
similar reduction was observed for 3-loop \textit{Arabidopsis}, with the 80
parameters describing the DE models decreasing to 20 in its Boolean
equivalent (Supp. Table S1). The ability of simple, discrete models to simulate biologically realistic
photoentraiment may also have important implications for the nature of
circadian signal processing, in addition to  being interesting from a modelling perspective. Specifically, it is consistent with hypotheses
based on experimental data suggesting that the mechanism by which daylength
and light intensity information is transmitted to output pathways may be
partly digital in nature \citep{Dodd05b}.

It should be noted that despite the success of the logic models in reproducing light responses, the coarser representation of network dynamics they provide means that the reproduction of certain circadian properties pose problems for the Boolean approach. A particular restriction of note relates to the modelling of temperature compensation. Temperature can be readily introduced into any DE model by following the methodology originally introduced by Ruoff \citep{Ruoff92,Ruoff94}. In this scheme, the temperature dependence of a kinetic parameter is assumed to be governed by the Arrhenius relation. This leads to a simple balance equation for the corresponding activation energies and control coefficients which, when satisfied, guarantees a near-zero period-temperature derivative at the balance point. In a Boolean model, the parameters that determine the free-running period $\tau_{FR}$ are the discrete delays $\tau_{j}$, and so simple balance equations can be derived from the form of the function determining the dependence of $\tau_{FR}$ on the $\tau_j$s. For example, in the 2-loop \textit{Neurospora} model, it can be shown that $\tau_{FR}=\tau_1+\tau_2+\tau_3+\tau_4$. Thus, compensation can be achieved simply by choosing the temperature dependence of each delay $\tau_j$ so that their sum is constant at each point over the range of interest.  However, as the $\tau_{j}$s are generic parameters which in each instance summarise several biochemical processes, any balance equation derived in this manner will not in general be grounded in physical chemistry, reducing its biological relevance.

\subsection{Computational advantages of the Boolean formulation}

In addition to providing a more compact parametrisation, a significant
strength of the logic modelling approach is that it greatly reduces the complexity of the optimisation procedure
itself. This enables optimal configurations and delay-threshold combinations to be
distinguished with a greater objectivity. The critical simplification afforded by the Boolean formulation is that the cost score is
computed from bitstrings, which take the value 0 or 1, rather than data
that varies over a much broader range of values. This removes the need for 
\textit{ad hoc} cost function terms for normalising the data and robustly
computing period and phase.  Indeed, the cost function we employed in this
work was extremely simple, computing the correlation between the predictions
of the model and the corresponding discretised experimental data at each
vertex.

A second important simplification relates to the structure of the parameter
space. In DE models, each interaction delay is dependent on processes that
occur over a range of time scales (e.g. transcription, translation,
degradation etc). This means that as well as being uncountably infinite, the
parameter space can cover several different orders of magnitude, making it
difficult to establish \textit{a priori} bounds. By contrast, each
signalling delay in a logic model is constrained to be a multiple of the
data sampling interval $t_{S}$, and is bounded above by the maximum possible
time $t_{MAX}$ over which the corresponding interaction can occur. $t_{MAX}$
is itself bounded by the duration of the experimental measurements and can
be further restricted on the basis of biological knowledge (here, the
free-running period was used to bound all delays - see the Methods section
for details). The set of possible signalling delays is thus finite. The set
of possible discretisation thresholds is also finite as varying the
threshold for a given species will generate a set number of different binary
time series, meaning it is only necessary to consider thresholds for which
distinct bitstrings are obtained.

For logic models, the parameter space as a whole is therefore finite, and
can be objectively bounded. Furthermore, for a fixed abstract topology, the
set of all possible logic configurations is also finite; it is simply equal
to the set of possible Boolean functions. This
means that it is, in principle, possible to comprehensively search across 
\textit{all} possible regulatory structures and parameter combinations to
determine the best fit to data. Such a search is impossible with DE systems. Furthermore, searching across different patterns of activation and inhibition for a DE model is often problematic as it requires some \textit{a priori} assumptions to be made regarding the underlying biochemical mechanisms; e.g. specification of which reactions may be cooperative and the ranges of the corresponding Hill coefficients \citep{Locke05a,Locke05b,Locke06,Akman08,Akman10}. In practice, however, the number of possible network and parameter combinations in a Boolean formulation can become too large for a complete search with the computational resources available, making it necessary to constrain the optimisation.

\subsection{Refining the optimisation protocol}

Here, in order to ensure computational tractability, we subsampled the delay-threshold space, while also
fixing the parameters controlling the impulse light inputs. In addition, we
restricted ourselves to a subset of logic circuits by assuming that: (i)
multiple light inputs to a gene are combined with an OR gate; and (ii) the
net light signal directly modulates the state of the gene through either an
AND or an OR\ gate, depending on the free-running light regime (see
Fig. \ref{fig:LCDiags}). Of course not all possible circuits will be
biologically reasonable ones (e.g. any circuit for which gates with light
inputs uniformly output 0 or 1 would be unviable). Nonetheless, it is
reasonable to expect that some interactions may be more accurately modelled
by gates that are not of the simple AND/OR type \citep{Buchler03}, meaning
that better performing logic configurations may have been overlooked.

In view of these restrictions, the fact that our optimal Boolean circuits
match the dynamics of their target data sets, both synthetic and
experimental, is very encouraging. Indeed, we anticipate that it should be
possible to find circuit configurations and parameter sets giving good fits
to data over a broader range of genetic and environmental perturbations. For
example, our Boolean 3-loop \textit{Arabidopsis} model shows consistency
with much of the photoperiodic behaviour of its DE counterpart (Fig. \ref{fig:PPs}D). However, it does not reproduce the phase response observed in the DE model as photoperiod is decreased, for which some components switch
between dawn- and dusk-dominance in a complex manner \citep{Edwards10}. 

A likely contributing factor is that our current optimisation method involves computing the score at every parameter
combination over a fixed lattice. This can be inefficient, particularly
where the stoichiometry of an interaction and/or its molecular dynamics
require the density of interacting species to accrue beyond a certain value
before the interaction is statistically likely to occur. In such cases, the
optimal threshold choice for the discretisation is likely be found within a
narrow band of possibilities. Furthermore, high threshold resolutions can be required to resolve topological degeneracies in the cost function which make optimal networks difficult to distinguish (see section S1.2 of the Electronic Supplementary Material). The accuracy of the optimisation could
thus be increased by employing logic variants of global search methods capable of
providing a more computationally efficient exploration of the parameter
space, such as evolutionary algorithms \citep{Mendes98}. This would increase
the predictive power of the models, better positioning us to address
features such as the dawn and dusk switching in shorter photoperiods observed in the 3-loop model.

\subsection{Future directions}

Finally, we note that there are many promising avenues for further developing the approaches
introduced in this study.  From a
theoretical perspective, there is scope for extending established methods for
analysing Boolean models. Of particular interest are techniques for determining the simplest inequalities involving linear combinations of the
time delays that are required to traverse a given path through the state
space \citep{Thomas83,Kaufman85,Kaufman99}. In the context of circadian
models, this would involve developing general methods for deriving linear inequalities that result in free-running cycles with the target period. As the computational demands of optimisation grow exponentially with the number of parameters to be fitted, restricting the search to the corresponding solution set would be expected to dramatically reduce the computational load.

More generally, hybrid logic/DE algorithms would see Boolean methods used firstly to
determine the optimal model configuration from data, and then to identify the
regions of parameter space within which an equivalent DE model is likely to
give a good fit. This would exploit the ability of logic models,
demonstrated in this work, to generate a coarse, but quantitative
representation of the system dynamics in a systematic and efficient manner. We anticipate that regulatory networks of much greater complexity than those considered here could be  quantitively modelled using such an approach.

\section{Materials and methods}

\subsection{Data sets used for model fitting}

In order to incorporate some of the variability in expression levels characteristic of real experimental data, synthetic data was generated using the variant of Gillespie's stochastic
simulation algorithm (SSA) introduced by Gonze \textit{et al.}  \citep{Gonze02a}. Fits to deterministic data obtained from direct integration of the DEs gave very similar results (data not shown). However, the stochastic data sets yielded a  clearer separation between LCs, most probably due to the lifting of topological degeneracies. For all circuits, final analyses were therefore restricted to the results obtained with stochastic time courses.

In generating the synthetic time courses, the
scaling (or extensitivity) parameter $\Omega$ was set close to the minimum
value yielding self-sustained, unforced oscillations in each case. The final
values used for simulations were: 1-loop \textit{Neurospora}, $\Omega =25;$
2-loop \textit{Neurospora}, $\Omega =50;$ 2- and 3-loop \textit{Arabidopsis}%
, $\Omega =1000$. Time series were generated for 5 cycles in both entrained
(LD 12:12) and free-running conditions (DD for the \textit{Neurospora}
models; LL for \textit{Arabidopsis}). This choice mirrors the combination of
experimental data sets typically chosen for parameter optimisation in
computational circadian studies %
\citep{Locke05a,Locke05b,Locke06,Akman08,Akman10,Pokhilko10}. Time series
were then subsampled every 0.5h to give the data used for model-fitting.
Plots of the synthetic LD data sets are shown in Supp. Fig. S5.

Experimental gene expression levels were measured using luciferase (LUC)
imaging, carried out as previously described \citep{Gould06}. Images were
recorded using Hamamatsu C4742-98 digital cameras operating at -75$^{\text{o}%
}$C under the control of Wasabi software (Hamamatsu Photonics, Hamamatsu
City, Japan) with a sampling interval of 1.5h. Bioluminescence levels were
quantified using Metamorph software (MDS, Toronto, Canada). The resulting
expression profiles were detrended for amplitude and baseline damping using
the mFourfit function of BRASSv3 \citep{Locke05b}. The detrended time series
can be seen in Fig. \ref{fig:expfreerunts}A.

\subsection{Implementing circadian networks in Boolean logic$\label%
{modmethodol}$}

For $n$ biochemical species, let $X_{i}\left( t\right)\in \left\{ 0,1\right\} $ denote
the activity of species $i$ at time $t$, where $t$ is a multiple, $kt_{S}$,
of the sampling interval $t_{S}$. The update rule is then expressed as:%
\begin{equation}
X_{i}\left( t\right) =s_{i}\left( X_{1}\left( t-\tau _{i_{1}}\right) ,\ldots
,X_{n}\left( t-\tau _{i_{n}}\right) ,L_{1}\left( t-\tau _{N+1}\right)
,\ldots ,L_{m}\left( t-\tau _{N+m}\right) \right) .  \label{eq:updaterule}
\end{equation}%
Here the $L_{k}$s represent external light inputs to the circuit and $\tau
_{1},\ldots ,\tau _{N+m}$ are the signalling delays, with $\tau _{i_{j}}\in
\left\{ \tau _{1},\ldots ,\tau _{N}\right\} $ for all $i,j$. Assuming a 24h
day and writing $t_{DAWN}$ and $t_{DUSK}$ for the times of dawn and dusk
respectively, the $L_{k}$s are described by the following function:%
\begin{equation}
L_{k}\left( t\right) =\left\{ 
\begin{array}{l}
1\text{ if }t_{DAWN}\leq \mathop{\rm mod}{\left( t,24\right)} \leq %
\mathop{\rm mod}{\left( t_{DAWN}+p_{k},24 \right)}, \\ 
0\text{ otherwise.}%
\end{array}%
\right.  \label{eq:lightfunc}
\end{equation}%
Setting $p_{k}=0$ creates a constant darkness signal (DD); setting $p_{k}=24$%
\ corresponds to constant light (LL); setting $p_{k}=t_{DUSK}-t_{DAWN}$
yields a continuous light-dark (LD) cycle. Parameter sets with $%
p_{k}<t_{DUSK}-t_{DAWN}$ yield LD\ cycles with a pulse of length $p_{k}$ at
dawn.

The functions $s_{i}$ are Boolean functions $s_{i}:\left\{ 0,1\right\}
^{n+m}\rightarrow \left\{ 0,1\right\} $ representing the interactions
between genes and light inputs that determine the state of species $X_{i}$.
We introduce a \textit{logical dependency} function (or \textit{logic gate}) 
$G\left( \cdot ,g\right) $ to describe these interactions. This function
takes a single parameter $g\in \left\{ 0,1\right\} $, the value of which
determines the type of reaction modelled. Formally, we consider two types of
operator. The first operator acts on a single Boolean input $Y\in \left\{
0,1\right\} $, implementing either the identity or NOT gate, modelling
activation and repression respectively: 
\begin{equation}
\begin{array}{l}
G_{1}\left( Y,0\right) =Y \\ 
G_{1}\left( Y,1\right) =\text{{\footnotesize NOT }}Y.%
\end{array}
\label{eq:gatetype1}
\end{equation}%
The second type of operator acts on two Boolean inputs $Y,Z\in \left\{
0,1\right\} $, implementing either the AND\ or the (inclusive) OR\
dependency. If either species can fulfil the interaction, an OR dependency
is used. If both species are required in the interaction, an AND dependency
is used. Thus, for species $Y\in \{0,1\}$ and $Z\in \{0,1\}$: 
\begin{equation}
\begin{array}{l}
G_{2}\left( Y,Z,0\right) =Y\text{{\footnotesize OR }}Z \\ 
G_{2}\left( Y,Z,1\right) =Y\text{{\footnotesize AND }}Z.%
\end{array}
\label{eq:gatetype2}
\end{equation}

The functions $s_{i}$ in (\ref{eq:updaterule}) are formed as compositions of these dependencies. For
example, the update rule for gene $TOC1$ in the optimal 2-loop \textit{%
Arabidopsis} model has the form 
\begin{equation}
X_{TOC1}\left( t\right) =\left( \text{{\footnotesize NOT }}X_{LHY}\left(
t-\tau _{1}\right) \right) \text{{\footnotesize AND }}X_{Y}\left( t-\tau
_{6}\right). \label{eq:arabidexode}
\end{equation}%
Using (\ref{eq:gatetype1}) and (\ref{eq:gatetype2}), this can be written
as a composition of logic functions%
\begin{equation}
X_{TOC1}\left( t\right) =G_{2}\left( G_{1}\left( X_{LHY}\left( t-\tau
_{1}\right) ,g_{1}\right) ,G_{1}\left( X_{Y}\left( t-\tau _{6}\right)
,g_{6}\right) ,g_{8}\right) ,  \label{eq:arabidex}
\end{equation}%
with $g_{6}=0$ and $g_{1}=g_{8}=1$. The resulting Boolean function models a
reaction in which $LHY$ must be downregulated in order for $Y$ to activate $%
TOC1$ production.

A given set of interaction functions $s_{i}$ has an associated adjacency
matrix $\mathbf{A}=\left( A_{ij}\right) $ defined by%
\begin{equation}
A_{ij}=\left\{ 
\begin{array}{l}
1\text{ if }s_{i}\left( X_{1},\ldots ,X_{j-1},0,X_{j+1}\ldots ,X_{n},\mathbf{L}%
\right) \notequiv s_{i}\left( X_{1},\ldots ,X_{j-1},1,X_{j+1}\ldots ,X_{n},%
\mathbf{L}\right), \\ 
0\text{ otherwise,}%
\end{array}%
\right.  \label{eq:Aij}
\end{equation}%
where $\mathbf{L}=\left( L_{1},\ldots ,L_{m}\right) $. Thus, $A_{ij}=1$ if
species $X_{j}$ can change the state of $X_{i}$, and so matrix $\mathbf{A}$
describes the abstract topology of a model (these topologies can be seen in
Fig. \ref{fig:LCDiags}). It follows that a circadian clock model is
parametrised by its adjacency matrix $\mathbf{A}$, a set of delays $\mathbf{%
\tau }=\left( \tau _{1},\ldots ,\tau _{N+m}\right) $ and a set of gates $%
\mathbf{G}=\left( g_{1},\ldots ,g_{d}\right) $. Writing $\mathbf{X}=\left(
X_{1},\ldots ,X_{n}\right) $, this yields the following compact, vectorised
form of the update rule:%
\begin{equation}
\mathbf{X}\left( t\right) =\mathbf{S}\left( \mathbf{X}\left( t\right) 
\mathbf{,L}\left( t\right) ;\mathbf{A,\tau ,G}\right).
\label{eq:updaterulevec}
\end{equation}%
We refer to the set of gates $\mathbf{G}$ as the model's \textit{logic
configuration} (LC). As each configuration $\mathbf{G}=\left( g_{1},\ldots
,g_{d}\right) $,\ is a bitstring, it can be represented uniquely by the
corresponding decimal expansion $D\left( \mathbf{G}\right) $ defined below:%
\begin{equation}
D\left( g_{1},\ldots ,g_{d}\right) =\sum_{l=1}^{d}g_{l}2^{d-l}.
\label{eq:decexp}
\end{equation}%
For the models considered, the LCs are enumerated in terms of their decimal
expansions for simplicity in Figs. \ref{fig:NeuroScores}, \ref{fig:ArabScores} and \ref{fig:ArabExpScores}.

Of the LCs consistent with $\mathbf{A}$, a subset matches the pattern of
activation and inhibition in the corresponding DE\ model. There are referred to as the \textit{DE\ LCs}. For example, for the 2-loop \textit{Neurospora}
model, \textit{frq} activates both isoforms of FRQ, giving $g_{1}=g_{2}=0$,
and both isoforms repress transcription, giving $g_{3}=g_{4}=1$ (see Fig. \ref{fig:DECircuitDiags}B and Fig. \ref{fig:LCDiags}B). There are thus
a pair of DE\ LCs in this case, $00110$ and $00111$, depending on whether
both isoforms are required for repression ($g_{5}=0$, corresponding to OR)
or a single isoform is sufficient ($g_{5}=1$, corresponding to AND). These DE\
LCs are encoded by the integers 6 and 7 respectively.

Finally, in order to construct the simplest logic models consistent with the
general form of circadian DE systems, each Boolean function $s_{i}$ is
assumed to have the form 
\begin{equation}
s_{i}\left( \mathbf{X},\mathbf{L}\right) =H_{i}\left( s_{i}^{F}\left( 
\mathbf{X}\right) ,s_{i}^{L}\left( \mathbf{L}\right) \right) ,
\label{eq:sicirc}
\end{equation}%
such that: (i) $H_{i}:\left\{ 0,1\right\} ^{2}\rightarrow \left\{
0,1\right\} $ implements either the AND\ or OR gate; (ii) $s_{i}^{F}:\left\{
0,1\right\} ^{n}\rightarrow \left\{ 0,1\right\} $ encodes the structure of
the free-running system; and (iii) $s_{i}^{L}:\left\{ 0,1\right\}
^{m}\rightarrow \left\{ 0,1\right\} $ determines how multiple light inputs
are integrated, with $s_{i}^{L}\left( 0,\ldots ,0\right) =0$ and $%
s_{i}^{L}\left( 1,\ldots ,1\right) =1$. For consistency, it is therefore
necessary that setting each light input $L_{k}$ to the relevant constant
value $L$ recovers the free-running system ($L=0$ for DD; $L=1$ for LL).
This condition is equivalent to $H_{i}\left( s_{i}^{F}\left( \mathbf{X}%
\right) ,L\right) =s_{i}^{F}\left( \mathbf{X}\right) $. For the \textit{%
Neurospora} models, where the free-running condition is DD, the consistency
condition is achieved using an OR\ gate since $x$ {\footnotesize OR }$0=x$.
In the case of the \textit{Arabidopsis} circuits, where the free-running
condition is LL, the AND gate is appropriate as $x$ {\footnotesize AND }$1=x$. 

For both \textit{Arabidopsis} models, the multiple light inputs to Y are
combined using an OR gate: this is because in the corresponding DE models,
both inputs can independently upregulate transcription %
\citep{Locke05b,Locke06}. In addition, the 3-loop model only incorporates
the pulsed light input to the PRR gene since removing the continuous light
input from the DE\ system had a negligible effect on its photoperiodic
behaviour.

Full details of the logic formulation of each clock network are given in
section S2 of the Electronic Supplementary Material.

\subsection{Optimisation and constraints}

In order to identify the optimal combination of signalling delays $\mathbf{%
\tau }=(\tau_{j})$, thresholds $\mathbf{T}=(T_{i})$ and logic gates $\mathbf{G}=(g_{l})$ for a given
model and data set, we must introduce a cost function to be minimised. We
used a simple function based on a correlation between the predicted time
courses generated by the model and the corresponding data. Further details
can be found in section S1.1 of the Electronic Supplementary Material. Optimal LC-parameter combinations are given in Tables 1 and 2.

The most comprehensive strategy in seeking to minimise the cost function is
to systematically calculate the cost for all possible configurations and 
parametrisations of the model; that is all delays $0\leq \tau _{j}\leq
t_{MAX}$, where $t_{MAX}$ is the longest timescale considered in the system,
and all thresholds $0<T_{i}<1$. However, when the model becomes too complex
to permit a global analysis, it becomes necessary to introduce further
constraints to restrict the parametrisation to be considered. Such
constraints should be as objective as possible.

One viable, objective constraint is to limit all the signalling delays
around a closed loop so that they sum to no more than the period $\tau _{FR}$
of free-running oscillations in the target data set.

For 1-loop \textit{Neurospora, }this gives the single delay bound%
\begin{equation}
\tau _{1}+\tau _{2}\leq \tau _{FR},  \label{eq:delbound1LNeuro}
\end{equation}%
where $\tau _{FR}=22$h.

For 2-loop \textit{Neurospora}, the constraint results in 2 delay bounds%
\begin{equation}
\begin{array}{c}
\tau _{1}+\tau _{3}\leq \tau _{FR}, \\ 
\tau _{2}+\tau _{4}\leq \tau _{FR},%
\end{array}
\label{eq:delbound2LNeuro}
\end{equation}%
where again $\tau _{FR}=22$h.

For 2-loop \textit{Arabidopsis}, there are 3 delay bounds:%
\begin{equation}
\begin{array}{r}
\tau _{1}+\tau _{2}+\tau _{3}\leq \tau _{FR}, \\ 
\tau _{5}+\tau _{6}\leq \tau _{FR}, \\ 
\tau _{2}+\tau _{3}+\tau _{4}+\tau _{6}\leq \tau _{FR}.%
\end{array}
\label{eq:delbound2LArabid}
\end{equation}%
In this case $\tau _{FR}=25$h.

For 3-loop \textit{Arabidopsis}, there are 4 delay bounds: those for the
2-loop model above together with an extra bound introduced by the addition
of the LHY-PRR\ loop:%
\begin{equation}
\tau _{7}+\tau _{8}\leq \tau _{FR}.  \label{eq:delbound3LArabid}
\end{equation}%
For this circuit, $\tau _{FR}=24$h.

Two further constraints were introduced for reasons of computational
tractability. Firstly, each delay $\tau _{j}$ was restricted to integer
multiples of a minimum delay resolution $\tau _{R}$, itself a multiple $k_{%
\mathbf{\tau }}t_{S}$ of the data sampling interval $t_{S}$. Secondly, each
threshold $T_{i}$ was bounded within a subinterval $\left[ T_{MIN},T_{MAX}%
\right] $ of $\left[ 0,1\right] $, and also restricted to integer multiples
of a minimum threshold resolution $T_{R}$.

\subsubsection{Optimisation to synthetic data}

For all models, $T_{MIN}$ and $T_{MAX}$ were set to the values $0.2$ and $%
0.8 $ respectively. For the \textit{Neurospora} models, $k_{\mathbf{\tau }}$
was set to $1$ and $T_{R}$ to $0.025$; for the \textit{Arabidopsis }models, $%
k_{\mathbf{\tau }}$ was initially set to $2$ and $T_{R}$ to $0.05$. Scores
were then recalculated with $k_{\mathbf{\tau }}=1$ and $T_{R}=0.025$, within
intervals $\left[ \tau _{i}-3\tau _{R},\tau _{i}+3\tau _{R}\right] $ and $%
\left[ T_{j}-5T_{R},T_{j}+5T_{R}\right] $\ centred around the parameter
combinations giving the best scores. For \textit{Arabidopsis} optimisations,
light parameters controlling the impulse inputs ($L_{1}$ and $L_{3}$ in the
2-loop circuit; $L_{1},L_{3}$ and $L_{4}$ in the 3-loop circuit) were fixed
at the values shown in Table \ref{tab:optimparams}. These were determined
from discrete approximations to the corresponding continuous curves in the
DE models.

Each parameter set was then checked for a functioning circadian clock.
Firstly, the solution to the free-running model was generated under the
appropriate continuous light conditions, using the discretised data as
initial conditions. If a limit cycle was obtained with period within $20\%$
of the DE\ model, this was fed into the model under simulated
12:12 LD cycles. Periodic, entrained oscillations were taken to indicate a viable clock circuit.

For the \textit{Neurospora} and 2-loop \textit{Arabidopsis} models,
optimisations included all possible LCs. The 3-loop \textit{Arabidopsis}
optimisations were restricted to the subset of LCs defined by $\mathbf{G}%
=\left( 10011011xyz\right) $, $x,y,z\in \left\{ 0,1\right\} $; these are the
configurations that result from fixing gates $g_{1},...,g_{7}$ to their
optimised values in the 2-loop circuit.

The photoperiod simulations shown in Fig. \ref{fig:PPs} were obtained by
locally reoptimising the logic circuits yielding viable clocks to simulations of the
DE models under 12:12 LD cycles, and then calculating the maximum symmetric
photoperiod interval $\left( 12-P_{MAX},12+P_{MAX}\right) $ over which both
model formulations generated stable, entrained solutions. The cost was
minimised with a simulated annealing algorithm %
\citep{Kirkpatrick83,Locke05a,Akman08}, using a cost function for which the
data and predicted time courses were taken to be single cycles of the
entrained solution in the DE\ and logic models respectively.

\subsubsection{Optimisation to experimental LUC data}

In fitting the 3-loop \textit{Arabidopsis} model to the LUC data shown in
Fig. \ref{fig:expfreerunts}A, genes were matched to model
variables in the following manner: (i) CCA1 was equated to LHY on the basis
that the LHY variable in the equivalent DE model groups together the effect
of the two genes \citep{Locke06}; (ii) TOC1 was matched to TOC1; (iii) GI\
was matched to Y owing to experimental results showing that this gene can
account for some of the action of Y in the DE model \citep{Locke06}; (iv)
PRR9 was matched to PRR, as this variable combines the PRR7 and 9 genes in
the DE formulation \citep{Locke06}.

Since the exact biological correlate of variable X is currently unknown, it
was not used for costing the fit. Consequently, $g_{2}$ and $\tau _{2}$ were
both fixed at 0 (note that fixing $g_{2}$ reduces the number of possible LCs
from 2048 to 1024). This preserves the dynamics of all components in the model excluding X, together with the
delay bounds used for constraining the parameter space. It also means that TOC1 and the dummy variable X have identical time series. Thus, in practice, the discretised TOC1 expression time series was used as a proxy for X data when calculating the
cost at the LHY\ vertex. Also, as the LUC data was measured in
LL, the parameters $p_{1}\rightarrow p_{4}$ controlling the light inputs
were all set to 24 (cf. eqn. (\ref{eq:lightfunc})).  In LL conditions, the absence of dawn and dusk means that the choice of the delay parameters $\tau _{9} \rightarrow \tau _{12}$ is arbitrary.  We therefore set the values of these delays to 0. $T_{MIN}$ and 
$T_{MAX}$ were fixed at $0.2$ and $0.8$ respectively. $k_{\mathbf{\tau }}$
was initially set to $2$ and $T_{R}$ to $0.2$. Scores were then recalculated
with $k_{\mathbf{\tau }}=1$ and $T_{R}=0.05$, within intervals $\left[ \tau
_{i}-2\tau _{R},\tau _{i}+2\tau _{R}\right] $ and $\left[
T_{j}-2T_{R},T_{j}+2T_{R}\right] $\ centred around the best scoring
parameter sets. The optimal parameter set for each LC was assessed to
determine whether it yielded a viable clock by first generating the solution
to the model using the discretised LUC time series as initial conditions,
and then checking that this gave a limit cycle with free-running period
within $20\%$ of 24h.

\subsection{Software}

The numerical routines for parameter optimisation and model simulation were
initially developed in MATLAB (Mathworks, Cambridge) and C. The scoring
algorithms used for global parameter sweeps were subsequently converted into Java and run on a task farm computer consisting of 118 Intel Harpertown
quad-core processors. All software used is available on request from the
corresponding author.

\section*{Acknowledgments}

The authors would like to thank Kieron Edwards for making available the
luciferase data used for model fitting and Declan Bates and Jonathan Fieldsend for helpful comments on the manuscript. CSBE is a Centre for Integrative
Systems Biology funded by BBSRC and EPSRC award D019621. This work was supported by  Wellcome Trust program grant WT066784 (to PG), and has made
use of resources provided by the Edinburgh Compute and Data Facility
(http://www.ecdf.ed.ac.uk/) through the eDIKT initiative ( http://www.edikt.org.uk).  AP\ was funded by a BBSRC\
Summer Studentship and a Moray Endowment small grant awarded by the
University of Edinburgh. The authors would also like to acknowledge the help of Konrad Paszkiewicz and Robin Batten in accessing HPC facilities provided by the University of Exeter under its Systems Biology initiative.

\clearpage

\section*{Figure legends}

\noindent
\textbf{Figure 1. Circuit diagrams for the clock models.} Genes are boxed and arrows
denote regulatory interactions. Diamonds represent light inputs. \textbf{A}.
The single-loop \textit{Neurospora} model \citep{Leloup99}. FRQ protein
represses production of \textit{frq} transcript. Light acts on the network
by upregulating \textit{frq} transcription. \textbf{B}. The two-loop \textit{%
Neurospora} model \citep{Akman08}. Two isoforms of FRQ are produced which
both repress \textit{frq} transcription. Light upregulates \textit{frq} as
in A. \textbf{C}. The two-loop \textit{Arabidopsis} model \citep{Locke05b}.
TOC1 activates its repressor LHY (combining LHY and CCA1) indirectly through
a hypothetical gene X, forming the central negative feedback loop of the
circuit. LHY is directly upregulated by light while light indirectly
activates TOC1 via a second hypothetical gene Y, posited to have two
distinct light inputs. Y activates TOC1 transcription and TOC1 represses Y, forming a second, interlocked feedback loop. \textbf{D}. The
three-loop \textit{Arabidopsis} model \citep{Locke06}. The additional PRR
gene (combining PRR7 and PRR9) is light-activated and represses LHY
transcription. LHY upregulates PRR, creating a third feedback loop. 
\vspace{-1.2cm}
\begin{figure}[h!]
\caption{}
\label{fig:DECircuitDiags}
\end{figure}

\noindent
\textbf{Figure 2. The abstract topologies for the logic representations of the clock circuits shown in Fig. \ref{fig:DECircuitDiags}.} Genes are boxed and arrows denote
regulatory interactions. \textbf{A}. The single-loop \textit{Neurospora} model. \textbf{B}. The two-loop \textit{Neurospora} model. \textbf{C}. The two-loop \textit{Arabidopsis} model. 
\textbf{D}. The three-loop \textit{Arabidopsis} model. Numerals $l$ index logic gates $g_{l}\in \left\{
0,1\right\}$ that can be varied to generate different regulatory structures.
Numerals at the end of an arrow index the single-input gate defining that
interaction. Numerals within boxes index logic gates governing double-input
interactions. Diamonds represent light inputs, with the corresponding fixed
gates in ovals indicating how these affect the target species (e.g. in 
C, $L_{2}$ and $L_{3}$ are combined with an OR gate after which the
resulting bitstring is combined with the output of Y through an AND gate -
see the Methods sections for full details). $\tau%
_j $s represent the circuit delays and $T_i $s the discretisation thresholds
used to fit continuous data.
\vspace{-1.2cm}
\begin{figure}[h!]
\caption{}
\label{fig:LCDiags}
\end{figure}

\noindent
\textbf{Figure 3. The number of parameters required for each clock configuration as
DE and logic models.} 
\vspace{-1.2cm}
\begin{figure}[h!]
\caption{}
\label{fig:hist_param}
\end{figure}

\noindent
\textbf{Figure 4. The results of exploring the logic configurations belonging to the
abstract topologies of the 1-loop (\textbf{A}) and 2-loop (\textbf{B}) 
\textit{Neurospora} models.} Cost scores are shown for the optimal fit of each LC to synthetic data.
The LCs are indexed by their decimal representations for brevity (see Methods for details).
Here, a score of 0 indicates the best fit and a score of 1 the worst fit.
Triangles indicate LCs for which the Boolean model yields a viable clock.
LCs mirroring the activation and inhibition pattern of the corresponding DE
models in Figs. \ref{fig:DECircuitDiags}A and B are plotted in red.
In A, one such LC mirrors the corresponding DE model, $G=(01)$, and
this emerges as the optimal configuration yielding a viable clock. In 
B, only LCs yielding scores less than 0.75 are shown. There are two
that mirror the equivalent DE model. One of these, $G=(00111)$, is
identified as the optimal configuration giving a viable clock (leftmost red
triangle). In this LC, either of the FRQ isoforms can independently inhibit
transcription (the corresponding two-input gate is of the AND type). 
\vspace{-1.2cm}
\begin{figure}[h!]
\caption{}
\label{fig:NeuroScores}
\end{figure}

\noindent
\textbf{Figure 5. The results of exploring the logic configurations belonging to the
abstract topologies of the 2-loop (\textbf{A}) and 3-loop (\textbf{B}) 
\textit{Arabidopsis} models.} Cost scores are shown for the optimal fit of each LC to synthetic data.
As in Fig. \ref{fig:NeuroScores}, each LC is indexed by its decimal
expansion. Scores of 0 and 1 indicate the best and worst fits respectively.
Triangles denote LCs for which the Boolean model yields a viable clock. LCs
mirroring the activation and inhibition pattern of the corresponding DE
models in Figs. \ref{fig:DECircuitDiags}C and D are plotted in red.
In A, only LCs yielding scores less than 0.75 are shown. Of these,
there are 3 LCs consistent with the activation and inhibition pattern of the
equivalent DE model from a possible total of 4. From this subset, $%
G=(10011011)$ emerges as the optimal configuration yielding a viable clock.
For this circuit, both LHY and TOC1 can independently inhibit Y while TOC1
is repressed unless LHY is repressed while Y is activated (the corresponding
gates are both of the AND type). In B, only the 8 LCs obtained by
varying the gates of the PRR-LHY loop were considered: all other gates were
fixed to those of the optimal 2-loop configuration. Of these 8
possible circuits, two LCs were consistent with the equivalent DE model,
from which $G=(10011011011)$ is identified as the optimal configuration
giving a viable clock. This LC corresponds to a clock network in which LHY
is repressed unless PRR is inactive and X is active (the corresponding gate
is of the AND type). 
\vspace{-1.2cm}
\begin{figure}[h!]
\caption{}
\label{fig:ArabScores}
\end{figure}

\noindent
\textbf{Figure 6. Time series for the differential equation and Boolean versions of
the clock models in 12:12 LD cycles.} Two 24h cycles are plotted for each
model. \textbf{A}, \textbf{B}: 1-loop \textit{Neurospora}; \textbf{C}, 
\textbf{D}: 2-loop \textit{Neurospora}; \textbf{E}, \textbf{F}: 2-loop 
\textit{Arabidopsis}; \textbf{G}, \textbf{H}: 3-loop \textit{Arabidopsis}.
Differential equation time series (left panels) have been normalised to lie between 0 and 1 in order to
facilitate comparison with the Boolean simulations (right panels). Different components
within a model are slightly offset from one another so they can be
distinguished more easily. The time step used for solving the Boolean models
was 0.5h, equal to the data sampling interval. 
\vspace{-1.2cm}
\begin{figure}[h!]
\caption{}
\label{fig:LDts}
\end{figure}

\noindent
\textbf{Figure 7. Comparing the photoperiodic behaviour of the Boolean and DE
versions of each model.} For Boolean models, the phase of each species is
taken as the time within the LD cycle of the ON to OFF transition (downward
triangles). Analogously, the DE model phases are defined as the times at
which species decrease below the thresholds yielding the optimal fit of the
corresponding Boolean circuit to data (upward triangles). Shaded areas of
plots, darkness; open areas, light. \textbf{A}. 1-loop \textit{Neurospora}.
The phase-photoperiod profiles are coincident, indicating that the Boolean
model exactly reproduces the photoperiodic behaviour of its DE counterpart:
FRQ transcript and protein are both locked to dusk across the photoperiod
range. \textbf{B}. 2-loop \textit{Neurospora}. The phase plots are almost
exactly equal, except for shorter photoperiods where they differ by the data
sampling interval. As for A, all components are locked to dusk. 
\textbf{C}. 2-loop \textit{Arabidopsis}. The Boolean and DE models exhibit
very similar patterns of dawn- and dusk-locking across genes. The two Y
phase-photoperiod profiles reflect the double peak observed in this
component (Fig. \ref{fig:LDts}E) which gives rise to a
(dawn-locked) light-induced peak and a (dusk-locked) circadian peak 
\citep{Edwards10}. \textbf{D}. 3-loop \textit{Arabidopsis}. The phase plots
are similar, with all components predominately dawn-locked. As for 2-loop 
\textit{Arabidopsis}, the two Y profiles reflect the double peak observed
for this gene (Fig. \ref{fig:LDts}G). 
\vspace{-1.2cm}
\begin{figure}[h!]
\caption{}
\label{fig:PPs}
\end{figure}

\noindent
\textbf{Figure 8. Identifying the logic configurations of the 3-loop \textit{%
Arabidopsis} model giving the best fits to experimental data.} Plotting conventions are as described in
Figs. \ref{fig:NeuroScores} and \ref{fig:ArabScores}. 
\textbf{A} shows the top ranking LCs . The black arrow indicates the optimal
configuration yielding a viable clock, $G_{OPT}=(10101011011)$. For this circuit,  LHY and TOC1 repress each other,  in agreement with recent biochemical evidence \citep{Gendron12}. The red
arrow denotes the second highest ranking viable LC, $G_{DE}=(10011011011)$.
This LC matches the regulatory structure of the DE model, and was previously
identified as the configuration yielding the best fit to synthetic data (see
Fig. \ref{fig:ArabScores}B). \textbf{B} plots the top ranking LCs
obtained under the constraint that LHY represses TOC1 and TOC1 activates LHY
(i.e. $g_{1}=1$ and $g_{2}=g_{3}=0$; see Fig. \ref{fig:LCDiags}D).
Under this assumption, $G_{DE}$ emerges as the optimal
clock circuit. The circuit diagrams for $G_{OPT}$ and $G_{DE}$ can be seen in Supp. Fig. S4.
\vspace{-1.2cm}
\begin{figure}[h!]
\caption{}
\label{fig:ArabExpScores}
\end{figure}

\noindent
\textbf{Figure 9. Simulations generated by the optimal fits of the 3-loop \textit{%
Arabidopsis} model to experimental data.} \textbf{A}. Experimental expression profiles for the genes CCA1, TOC1, GI and PRR9 in free-running conditions (LL).
Expression levels were determined using LUC reporter gene imaging constructs
and have been normalised to lie between 0 and 1. \textbf{B}. The equivalent
Boolean time series generated by the logic configuration $G_{OPT}$ yielding the best fit to data. \textbf{C}. Boolean expression profiles for the highest ranked configuration $G_{DE}$ incorporating the central LHY-CCA1 negative feedback loop of the DE model. In all plots, different components are
slightly offset from one another so they can be distinguished more easily.
The time step used for solving the logic model was 1.5h, equal to the
data sampling interval.
\vspace{-1.2cm}
\begin{figure}[h!]
\caption{}
\label{fig:expfreerunts}
\end{figure}

\clearpage

\section*{Tables}
\begin{table}[!h]
\caption{
\bf{Optimal parameter sets - synthetic data.}}
\begin{center}
\begin{tabular}{|c||c|c|c|c|}
\hline
& One-loop & Two-loop & Two-loop & Three-loop \\ 
& \textit{Neurospora} & \textit{Neurospora} & \textit{Arabidopsis} & \textit{%
Arabidopsis} \\ \hline\hline
$\mathbf{G}$ & 01 & 00111 & 10011011 & 10011011011 \\ \hline\hline
$\tau_1$ (h) & 5 (5) & 5 (5.5) & 1.5 (3) & 0 (3) \\ \hline
$\tau_2$ (h) & 6.5 (6.5) & 1.5 (2) & 5.5 (6) & 5.5 (6.5) \\ \hline
$\tau_3$ (h) & 7.5 (9) & 6 (6.5) & 6.5 (7.5) & 7 (8) \\ \hline
$\tau_4$ (h) & - & 10 (9) & 0 (0.5) & 0 (1) \\ \hline
$\tau_5$ (h) & - & 9 (9) & 7.5 (6) & 8 (5) \\ \hline
$\tau_6$ (h) & - & - & 4 (4) & 5 (3.5) \\ \hline
$\tau_7$ (h) & - & - & 0 (1) & 4.5 (6) \\ \hline
$\tau_8$ (h) & - & - & 2.5 (0.5) & 6 (5) \\ \hline
$\tau_9$ (h) & - & - & 1 (0) & 1 (0) \\ \hline
$\tau_{10}$ (h) & - & - & - & 1 (3) \\ \hline
$\tau_{11}$ (h) & - & - & - & 0.5 (0) \\ \hline
$\tau_{12}$ (h) & - & - & - & 3 (4) \\ \hline\hline
\mbox{$T_1$} & 0.35 (0.40) & 0.425 (0.425) & 0.250 (0.450) & 0.1250 (0.4350) \\ \hline
\mbox{$T_2$} & 0.40 (0.70) & 0.525 (0.650) & 0.375 (0.775) & 0.3000 (0.6850) \\ \hline
\mbox{$T_3$} & - & 0.250 (0.350) & 0.575 (0.750) & 0.6250 (0.7325) \\ \hline
\mbox{$T_4$} & - & - & 0.150 (0.325) & 0.2250 (0.1295) \\ \hline
\mbox{$T_5$} & - & - & - & 0.8250 (0.9500) \\ \hline\hline
\mbox{$p_{1}$ (h)} & $P$ & $P$ & 2 & 2 \\ \hline
\mbox{$p_{2}$ (h)} & - & $P$ & $P$ & $P$ \\ \hline
\mbox{$p_{3}$ (h)} & - & - & 0.5 & 0.5 \\ \hline
\mbox{$p_{4}$ (h)} & - & - & - & 3 \\ \hline
\end{tabular}%
\end{center}
\begin{flushleft} The logic configurations, $\mathbf{G}$, delays, $\protect\tau _{j}$%
, and discretisation thresholds, $T_{i}$ ($0 <T_{i}<1$), yielding the
best fit of each logic model to synthetic time series. The values used for photoperiod simulations - obtained by fitting directly to DE time series - are shown in brackets. For each model, $p_{k}$
indicates the parameter used to simulate light input $L_{k}$ through
equation (\protect\ref{eq:lightfunc}). These were fixed at the values shown,
with $P$ denoting the photoperiod $t_{DUSK}-t_{DAWN}$.

\end{flushleft}
\label{tab:optimparams}
\end{table}

\begin{table}[h!]
\caption{
\bf{Optimal parameter sets - experimental data.}}
\begin{center}
\begin{tabular}{|c||c|c|}
\hline
$\mathbf{G}$ & 10101011011 ($G_{OPT}$) & 10011011011 ($G_{DE}$) \\ 
\hline\hline
$\tau_1$ (h) & 1.5 & 1.5 \\ \hline
$\tau_2$ (h) & 0 & 0 \\ \hline
$\tau_3$ (h) & 1.5 & 10.5 \\ \hline
$\tau_4$ (h) & 9 & 0 \\ \hline
$\tau_5$ (h) & 6 & 7.5 \\ \hline
$\tau_6$ (h) & 4.5 & 4.5 \\ \hline
$\tau_7$ (h) & 1.5 & 3 \\ \hline
$\tau_8$ (h) & 10.5 & 6 \\ \hline
$\tau_9$ (h) & 0 & 0 \\ \hline
$\tau_{10}$ (h) & 0 & 0 \\ \hline
$\tau_{11}$ (h) & 0 & 0 \\ \hline
$\tau_{12}$ (h) & 0 & 0 \\ \hline\hline
\mbox{$T_1$} & 0.30 & 0.35 \\ \hline
\mbox{$T_2$} & 0.40 & 0.35 \\ \hline
\mbox{$T_3$} & - & - \\ \hline
\mbox{$T_4$} & 0.15 & 0.10 \\ \hline
\mbox{$T_5$} & 0.50 & 0.45 \\ \hline\hline
\mbox{$p_{1}$ (h)} & 24 & 24 \\ \hline
\mbox{$p_{2}$ (h)} & 24 & 24 \\ \hline
\mbox{$p_{3}$ (h)} & 24 & 24 \\ \hline
\mbox{$p_{4}$ (h)} & 24 & 24 \\ \hline
\end{tabular}%
\end{center}
\begin{flushleft} The logic configurations, $\mathbf{G}$, delays, $\protect\tau _{j}$%
, and discretisation thresholds, $T_{i}$ ($0 < T_{i} < 1$), yielding the
top two fits of the 3-loop \textit{Arabidopsis} logic model to experimental
LUC time series recorded in LL. $G_{OPT}$ is the highest-scoring LC. $G_{DE}$ is
the second highest scoring LC, and is also the top ranked configuration
under the constraint that LHY represses TOC1 and TOC1 promotes LHY
production. Note that $G_{DE}$ was previously identified as the LC giving
the optimal fit to synthetic data generated by the equivalent DE model (see
Table \protect\ref{tab:optimparams}).
\end{flushleft}
\label{tab:optimparamsexp}
\end{table}


\newpage

\begin{thebibliography}{10}

\bibitem{Dunlap03}
Dunlap, J.~C., Loros, J.~J. \& DeCoursey, P.~J.
\newblock 2003 {\em Chronobiology: Biological Timekeeping}.
\newblock (Sinauer).

\bibitem{Young01}
Young, M.~W. \&  Kay, S.~A.
\newblock 2001 Time zones: a comparative genetics of circadian clocks.
\newblock {\em Nat. Rev. Genet.} {\bf 2}, 702--15.

\bibitem{Bell-Pedersen05}
Bell-Pedersen, D., Cassone, V.~M., Earnest, D.~J., Golden, S.~S., Hardin,
  P.~E., Thomas, T.~L. \& Zoran, M.
\newblock 2005 Circadian rhythms from multiple oscillators: lessons from
  diverse organisms.
\newblock {\em Nat. Rev. Genet.} {\bf 6}, 544--556.

\bibitem{Zhang10}
Zhang, E.~E. \& Kay, S.~A.
\newblock 2010 Clocks not winding down: unravelling circadian networks.
\newblock {\em Nat. Rev. Mol. Cell Biol.} {\bf 11}, 764--776.

\bibitem{Khapre10}
Khapre, R.~V., Samsa, W.~E. \& Kondatov, R.~V.
\newblock 2010 Circadian regulation of cell cycle: {Molecular} connections
  between aging and the circadian clock.
\newblock {\em Ann. Med.} {\bf 42}, 1695--1700.

\bibitem{Gery10}
Gery, S. \& Koeffler, H.~P.
\newblock 2010 Circadian rhythms and cancer.
\newblock {\em Cell Cycle} {\bf 9}, 1097--1103.

\bibitem{Takeda10}
Takeda, N. \& Maemura., K.
\newblock 2010 Circadian clock and vascular disease.
\newblock {\em Hypertens. Res.} {\bf 33}, 645--651.

\bibitem{Westrich10}
Westrich, L. \& Sprouse., J.
\newblock 2010 Circadian rhythm dysregulation in bipolar disorder.
\newblock {\em Curr. Opin. Investig. Drugs} {\bf 11}, 779--787.

\bibitem{Lange06}
Lange, T., Dimitrov, S., Fehm, H.~L., Westermann, J. \& Born, J.
\newblock 2006 Shift of monocyte function toward cellular immunity during
  sleep.
\newblock {\em Arch. Intern. Med.} {\bf 166}, 1695--1700.

\bibitem{Liu06}
Liu, J., Malkani, G., Mankani, G., Shi, X., Meyer, M., Cunningham-Runddles, S. \& Sun, Z.~S.
\newblock 2006 The circadian clock {Period 2} gene regulates gamma interferon
  production of {NK} cells in host response to lipopolysaccharide-induced
  endotoxic shock.
\newblock {\em Infect. Immun.} {\bf 74}, 4750--4756.

\bibitem{Keller09}
Keller, M., Mazuch, J., Abraham, U., Eom, G.~D., Herzog, E.~D., Volk, H.~D.,
  Kramer, A. \& Maier, B.
\newblock 2009 A circadian clock in macrophages controls inflammatory immune
  responses.
\newblock {\em Proc. Natl. Acad. Sci. USA} {\bf 106}, 21407--12.

\bibitem{Ouyang98}
Ouyang, Y., Andersson, C.~R., Kondo, T., Golden, S.~S. \& Johnson, C.~H.
\newblock 1998 Resonating circadian clocks enhance fitness in cyanobacteria.
\newblock {\em Proc. Natl. Acad. Sci. USA} {\bf 95}, 8660--4.

\bibitem{Dodd05a}
Dodd, A.~N., Salathia, N., Hall, A., K{\'e}vei, E., T{\'o}th, R., Nagy, F.,
  Hibberd, J.~M., Millar, A.~J. \& Webb, A.~A.
\newblock 2005 Plant circadian clocks increase photosynthesis, growth,
  survival, and competitive advantage.
\newblock {\em Science} {\bf 309}, 630--3.

\bibitem{Gardner81}
Gardner, G.~F. \& Feldman, J.~F.
\newblock 1981 Temperature compensation of circadian period length in clock
  mutants of \textit{Neurospora crassa}.
\newblock {\em Plant Physiol.} {\bf 68}, 1244--1248.

\bibitem{Ruoff97}
Ruoff, P., Rensing, L., Kommedal, R. \& Mohsenzadeh, S.
\newblock 1997 Modeling temperature compensation in chemical and biological
  oscillators.
\newblock {\em Chronobiol. Int.} {\bf 14}, 499--510.

\bibitem{Gould06}
Gould, P.~D., Locke, J.~C., Larue, C., Southern, M.~M., Davis, S.~J., Hanano,
  S., Moyle, R., Milich, R., Putterill, J., Millar, A.~J. \& Hall, A.
\newblock 2006 The molecular basis of temperature compensation in the
  \textit{Arabidopsis} circadian clock.
\newblock {\em Plant Cell} {\bf 18}, 1177--1187.

\bibitem{Brunner06}
Brunner, M. \& Diernfellner, A.
\newblock 2006 How temperature affects the circadian clock of \textit{Neurospora
  crassa}.
\newblock {\em Chronobiol. Int.} {\bf 23}, 81--90.

\bibitem{Akman08}
Akman, O.~E., Locke, J. C.~W., Tang, S., Carr{\'e}, I., Millar, A.~J. \& Rand, D.~A.
\newblock 2008 Isoform switching facilitates period control in the
  \textit{Neurospora crassa} circadian clock.
\newblock {\em Mol. Syst. Biol.} {\bf 4}, 64.

\bibitem{Leloup99}
Leloup, J.-C., Gonze, D. \& Goldbeter, A.
\newblock 1999 Limit cycle models for circadian rhythms based on
  transcriptional regulation in \textit{Drosophila} and \textit{Neurospora}.
\newblock {\em J. Biol. Rhythms} {\bf 14}, 433--448.

\bibitem{Smolen01}
Smolen, P., Baxter, D.~A. \& Byrne, J.~H.
\newblock 2001 Modeling circadian oscillations with interlocking positive and
  negative feedback loops.
\newblock {\em J. Neurosci.} {\bf 21}, 6644--56.

\bibitem{Sriram04}
Sriram, K. \& Gopinathan, M.~S.
\newblock 2004 A two variable delay model for the circadian rhythm of
  \textit{Neurospora crassa}.
\newblock {\em J. Theor. Biol.} {\bf 231}, 23--38.

\bibitem{Francois05}
Francois, P.
\newblock 2005 A model for the \textit{Neurospora} circadian clock.
\newblock {\em Biophys. J.} {\bf 88}, 2369--2383.

\bibitem{Ruoff05}
Ruoff, P., Loros, J.~J. \& Dunlap, J.~C.
\newblock 2005 The relationship between {FRQ}-protein stability and
  temperature compensation in the \textit{Neurospora} circadian clock.
\newblock {\em Proc. Natl. Acad. Sci. USA} {\bf 102}, 17681--6.

\bibitem{Hong08a}
Hong, C.~I., Jolma, I.~W., Loros, J.~J., Dunlap, J.~C. \& Ruoff, P.
\newblock 2008 Simulating dark expressions and interactions of frq and wc-1
  in the \textit{Neurospora} circadian clock.
\newblock {\em Biophys. J.} {\bf 94}, 1221--32.

\bibitem{Akman10}
Akman, O.~E., Rand, D.~A., Brown, P.~E. \& Millar, A.~J.
\newblock 2010 Robustness from flexibility in the fungal circadian clock.
\newblock {\em BMC Syst. Biol.} {\bf 4}, 88.

\bibitem{Goldbeter95}
Goldbeter, A.
\newblock 1995 A model for circadian oscillations in the \textit{Drosophila} period
  protein {PER}.
\newblock {\em Proc. Biol. Sci.} {\bf 261}, 319--24.

\bibitem{Leloup98}
Leloup, J.~C. \& Goldbeter, A.
\newblock 1998 A model for circadian rhythms in \textit{Drosophila} incorporating
  the formation of a complex between the {PER} and {TIM} proteins.
\newblock {\em J. Biol. Rhythms} {\bf 13}, 70--87.

\bibitem{Ueda01}
Ueda, H.~R., Hagiwara, M. \& Kitano, H.
\newblock 2001 Robust oscillations within the interlocked feedback model of
  \textit{Drosophila} circadian rhythm.
\newblock {\em J. Theor. Biol.} {\bf 210}, 401--6.

\bibitem{Smolen04}
Smolen, P., Hardin, P.~E., Lo, B.~S., Baxter, D.~A. \& Byrne, J.~H.
\newblock 2004 Simulation of \textit{Drosophila} circadian oscillations, mutations,
  and light responses by a model with {VRI}, {PDP-1}, and {CLK}.
\newblock {\em Biophys. J.} {\bf 86}, 2786--802.

\bibitem{Forger03}
Forger, D.~B. \& Peskin, C.~S.
\newblock 2003 A detailed predictive model of the mammalian circadian clock.
\newblock {\em Proc. Natl. Acad. Sci. USA} {\bf 100}, 14806--11.

\bibitem{Leloup03}
Leloup, J.~C. \& Goldbeter, A.
\newblock 2003 Toward a detailed computational model for the mammalian
  circadian clock.
\newblock {\em Proc. Natl. Acad. Sci. USA} {\bf 100}, 7051--6.

\bibitem{Becker04}
Becker-Weimann, S., Wolf, J., Herzel, H. \& Kramer, A.
\newblock 2004 Modeling feedback loops of the mammalian circadian oscillator.
\newblock {\em Biophys. J.} {\bf 87}, 3023--3034.

\bibitem{Mirsky09}
Mirsky, H.~P., Liu, A.~C., Welsh, D.~K., Kay, S.~A. \& {Doyle III}, F.~J.
\newblock 2009 A model of the cell-autonomous mammalian circadian clock.
\newblock {\em Proc. Natl. Acad. Sci. USA} {\bf 106}, 11107--12.

\bibitem{Locke05a}
Locke, J. C.~W., Millar, A.~J. \& Turner, M.~S.
\newblock 2005 Modelling genetic networks with noisy and varied experimental
  data: the circadian clock in \textit{Arabidopsis thaliana}.
\newblock {\em J. Theor. Biol.} {\bf 234}, 383--93.

\bibitem{Locke05b}
Locke, J. C.~W., Southern, M.~M., Kozma-Bognar, L., Hibberd, V., Brown, P.~E.,
  Turner, M.~S. \& Millar, A.~J.
\newblock 2005 Extension of a genetic network model by iterative
  experimentation and mathematical analysis.
\newblock {\em Mol. Syst. Biol.} {\bf 1}, 2005.0013.

\bibitem{Locke06}
Locke, J. C.~W., Kozma-Bognar, L., Gould, P.~D., Feh{\'e}r, B., Kevei, E.,
  Nagy, F., Turner, M.~S., Hall A. \& Millar, A.~J.
\newblock 2006 Experimental validation of a predicted feedback loop in the
  multi-oscillator clock of \textit{Arabidopsis thaliana}.
\newblock {\em Mol. Syst. Biol.} {\bf 2}, 59.

\bibitem{Zeilinger06}
Zeilinger, M.~N., Farr{\'e}, E.~M., Taylor, S.~R., Kay, S.~A. \& {Doyle III}, F.~J.
\newblock 2005 A novel computational model of the circadian clock in
  \textit{Arabidopsis} that incorporates prr7 and prr9.
\newblock {\em Mol. Syst. Biol.} {\bf 2}, 58.

\bibitem{Pokhilko10}
Pokhilko, A., Hodge, S.~K., Stratford, K., Knox, K., Edwards, K., Thomson,
  A.~W., Mizuno, T. \& Millar, A.~J.
\newblock 2010 Data assimilation constrains new connections and components in
  a complex, eukaryotic circadian clock model.
\newblock {\em Mol. Syst. Biol.} {\bf 6}, 416.

\bibitem{Gonze02a}
Gonze, D., Halloy, J. \& Goldbeter, A.
\newblock 2002 Biochemical clocks and molecular noise: Theoretical study of
  robustness factors.
\newblock {\em J. Chem. Phys.} {\bf 116}, 10997--11010.

\bibitem{Gonze02b}
Gonze, D., Halloy, J. \& Goldbeter, A.
\newblock 2002 Robustness of circadian rhythms with respect to molecular
  noise.
\newblock {\em Proc. Natl. Acad. Sci. USA} {\bf 99}, 673--8.

\bibitem{Vilar02}
Vilar, J.~M., Kueh, H.~Y., Barkai, N. \& Leibler, S.
\newblock 2002 Mechanisms of noise-resistance in genetic oscillators.
\newblock {\em Proc. Natl. Acad. Sci. USA} {\bf 99}, 5988--5992.

\bibitem{Kauffman69}
Kauffman, S.~A.
\newblock 1969 Metabolic stability and epigenesis in randomly constructed
  genetic nets.
\newblock {\em J. Theor. Biol.} {\bf 22}, 437--467.

\bibitem{Thomas91}
Thomas, R.
\newblock 1991 Regulatory networks seen as asynchronous automata: {A} logical
  description.
\newblock {\em J. Theor. Biol.} {\bf 153}, 1--23.

\bibitem{Kaufman99}
Kaufman, M., Andris, F., Leo, O.
\newblock 1999 A logical analysis of {T} cell activation and anergy.
\newblock {\em Proc. Natl. Acad. Sci. USA} {\bf 7}, 3894--3899.

\bibitem{Shmulevich02a}
Shmulevich, I., Dougherty, E.~R., Kim, S. \& Zhang, W.
\newblock 2002 Probabilistic {Boolean} networks: a rule-based uncertainty
  model for gene regulatory networks.
\newblock {\em Bioinformatics} {\bf 18}, 261--274.

\bibitem{Watterson08a}
Watterson, S., Marshall, S. \& Ghazal, P.
\newblock 2008 Logic models of pathway biology.
\newblock {\em Drug. Discov. Today} {\bf 13}, 447--456.

\bibitem{Watterson08b}
Yu, L., Watterson, S., Marshall, S. \& Ghazal, P.
\newblock 2008 Inferring {Boolean} networks with perturbation from sparse
  gene expression data: a general model applied to the {Interferon} regulatory
  network.
\newblock {\em Mol. Biosyst.} {\bf 4}, 1024--1030.

\bibitem{Saez09}
Saez-Rodriguez, J., Alexopoulos, L.~G., Epperlein, J., Samaga, R.,
  Lauffenburger, D.~A., Klamt, S. \& Sorger, P.~K.
\newblock 2009 Discrete logic modelling as a means to link protein signalling
  networks with functional analysis of mammalian signal transduction.
\newblock {\em Mol. Syst. Biol.} {\bf 5}, 331.

\bibitem{Schlatter09}
Schlatter, R., Schmich, K., {Avalos Vizcarra}, I., Scheurich, P., Sauter, T.,
  Borner, C., Ederer, M., Merfort, I. \& Sawodny, O.
\newblock 2009 {ON/OFF} and beyond - a {Boolean} model of apoptosis.
\newblock {\em PLoS Comput. Biol.} {\bf 5}, e1000595.

\bibitem{Watterson10}
Watterson, S. \& Ghazal, P.
\newblock 2010 Use of logic theory in understanding regulatory pathway
  signaling in response to infection.
\newblock {\em Future Microbiol.} {\bf 5}, 163--176.

\bibitem{Gendron12}
Gendron, J.~M., Pruneda-Paz, J.~L., Doherty, C.~J., Gross, A.~M., Kang, S.~E.
  \& Kay, S.~A.
\newblock 2012 \textit{Arabidopsis} circadian clock protein, {TOC1}, is a
  {DNA}-binding transcription factor.
\newblock {\em Proc. Natl. Acad. Sci. USA} {\bf 109}, 3167--72.

\bibitem{Thomas83}
Thomas, R.
\newblock 1983 Logical description, analysis and synthesis of biological and
  other networks comprising feedback loops.
\newblock {\em Adv. Chem. Phys.} {\bf 55}, 247--282.

\bibitem{Kaufman85}
Kaufman, M., Urbain, J. \& Thomas, R.
\newblock 1985 Towards a logical analysis of the immune response.
\newblock {\em J. Theor. Biol.} {\bf 114}, 527--561.

\bibitem{Shmulevich02b}
Shmulevich, I., Dougherty, E.~R. \& Zhang, W.
\newblock 2002 From {Boolean} to probabilistic {Boolean} networks as models
  of genetic regulatory networks.
\newblock {\em Proc. IEEE} {\bf 90}, 1778--1792.

\bibitem{Troein10}
Troein, C., Locke, J.~C., Turner, M.~S. \& Millar, A.~J.
\newblock 2009 Weather and seasons together demand complex biological clocks.
\newblock {\em Curr. Biol.} {\bf 19}, 1961--4.

\bibitem{Tan04}
Tan, Y., Dragovic, Z., Roenneberg, T. \& Merrow, M.
\newblock 2004 Entrainment dissociates transcription and translation of a
  circadian clock gene in \textit{Neurospora}.
\newblock {\em Curr. Biol.} {\bf 14}, 433--8.

\bibitem{Edwards10}
Edwards, K.~D., Akman, O.~E., Knox, K., Lumsden, P.~J., Thomson, A.~W., Brown,
  P.~E., Pokhilko, A., Kozma-Bognar, L., Nagy, F., Rand, D.~A. \& Millar,
  A.~J.
\newblock 2010 Quantitative analysis of regulatory flexibility under changing
  environmental conditions.
\newblock {\em Mol. Syst. Biol.} {\bf 6}, 424.

\bibitem{Ruoff96}
Ruoff, P. \& Rensing, L.
\newblock 1996 The temperature-compensated {Goodwin} model simulates many
  circadian clock properties.
\newblock {\em J. Theor. Biol.} {\bf 179}, 275--285.

\bibitem{Ay09}
Ay, F., Xu, F. \& Kahveci, T.
\newblock 2009 Scalable steady state analysis of {Boolean} biological
  regulatory networks.
\newblock {\em PLoS ONE} {\bf 4}, e7992.

\bibitem{Hau11}
Hau, L.~D. \& Kwon, Y.~K.
\newblock 2011 The effects of feedback loops on disease comorbidity in human
  signaling networks.
\newblock {\em Bioinformatics} {\bf 27}, 1113--20.

\bibitem{Krawitz07a}
Krawitz, P. \& Shmulevich, I.
\newblock 2007 Basin entropy in {Boolean} network ensembles.
\newblock {\em Phys. Rev. Lett.} {\bf 98}, 158701.

\bibitem{Ribeiro07}
Ribeiro, A.~S. \& Kauffman, S.~A.
\newblock 2007 Noisy attractors and ergodic sets in models of gene regulatory
  networks.
\newblock {\em J. Theor. Biol.} {\bf 247}, 743--55.

\bibitem{Garg08}
Garg, A., Cara, A.~D., Xenarios, I., Mendoza, L. \& Micheli, G.~D.
\newblock 2008 Synchronous versus asynchronous modeling of gene regulatory
  networks.
\newblock {\em Bioinformatics} {\bf 24}, 1917--25.

\bibitem{Buchler03}
Buchler, N., Gerland, U. \& Hwa, T.
\newblock 2003 On schemes of combinatorial transcription logic.
\newblock {\em Proc. Natl. Acad. Sci. USA} {\bf 100}, 5136--5141.

\bibitem{Gerstung09}
Gerstung, M., Timmer, J. \& Fleck, C.
\newblock 2009 Noisy signaling through promoter logic gates.
\newblock {\em Phys. Rev. E} {\bf 79}, 011923.

\bibitem{Hunziker10}
Hunziker, A., Tuboly, C., Horv{\'a}th, P., Krishna, S. \& Semsey, S.
\newblock 2010 Genetic flexibility of regulatory networks.
\newblock {\em Proc. Natl. Acad. Sci. USA} {\bf 107}, 12998--3003.

\bibitem{Milo02}
Milo, R., Shen-Orr, S., Itzkovitz, S., Kashtan, N., Chklovskii, D. \& Alon,
  U.
\newblock 2002 Network motifs: Simple building blocks of complex networks.
\newblock {\em Science} {\bf 298}, 824--827.

\bibitem{Prill05}
Prill, R., Iglesias, P. \& Levchenko, A.
\newblock 2005 Dynamic properties of network motifs contribute to biological
  network organization.
\newblock {\em PLoS Biol.} {\bf 3}, e343.

\bibitem{Alabadi01}
Alabad{\'\i}, D., Oyama, T., Yanovsky, M.~J., Harmon, F.~G., M{\'a}s, P. \& Kay, S.~A.
\newblock 2001 Reciprocal regulation between {TOC1} and {LHY/CCA1} within the
  \textit{Arabidopsis} circadian clock.
\newblock {\em Science} {\bf 293}, 880--883.

\bibitem{Pokhilko12}
Pokhilko, A., Fernandez, A.~P., Edwards, K.~D., Southern, M.~M., Halliday,
  K.~J. \& Millar, A.~J.
\newblock 2012 The clock gene circuit in \textit{Arabidopsis} includes a
  repressilator with additional feedback loops.
\newblock {\em Mol. Syst. Biol.} {\bf 8}, 574.

\bibitem{Dodd05b}
Dodd, A.~N., Love, J. \& Webb, A.~A.
\newblock 2005 The plant clock shows its metal: circadian regulation of
  cytosolic free {Ca$^{2+}$}.
\newblock {\em Trends Plant Sci.} {\bf 10}, 15--21.

\bibitem{Ruoff92}
Ruoff, P.
\newblock 1992 Introducing temperature-compensation in any reaction kinetic
  oscillator model.
\newblock {\em Interdiscipl. Cycle Res.} {\bf 23}, 92--99.

\bibitem{Ruoff94}
Ruoff, P.
\newblock 1994 General homeostasis in period and temperature-compensated
  chemical clock mutants by random selection conditions.
\newblock {\em Naturwissenschaften} {\bf 81}, 456---459.

\bibitem{Mendes98}
Mendes, P. \& Kell, D.
\newblock 1998 Non-linear optimization of biochemical pathways: applications
  to metabolic engineering and parameter estimation.
\newblock {\em Bioinformatics} {\bf 14}, 869--83.

\bibitem{Kirkpatrick83}
Kirkpatrick, S., Gelatt, C.~D. \& Vecchi, M.~P.
\newblock 1983 Optimization by simulated annealing.
\newblock {\em Science} {\bf 220}, 671--680.

\end{thebibliography}
\end{document}